\documentclass[aps,showpacs,nofootinbib]{revtex4}
\usepackage{subfig}
\usepackage{float}
\usepackage{epsfig}
\usepackage{graphicx}
\usepackage{graphics,epsfig}
\usepackage{amsmath}
\usepackage{amssymb}

\usepackage[usenames,dvipsnames]{color}
\newcommand{\be}{\begin{equation}}
\newcommand{\ee}{\end{equation}}
\newcommand{\ba}{\begin{eqnarray}}
\newcommand{\ea}{\end{eqnarray}}

\newcommand{\bwt}{ \begin{widetext}}
\newcommand{\ewt}{ \end{widetext}}

\newcommand{\beq}{\begin{equation}}
\newcommand{\eeq}{\end{equation}}

\newcommand{\half}{\frac 1 2 }

\begin{document}
\title{A Born-Infeld-like $f(R)$ gravity}
\author{\ J.C. Fabris}
\email{fabris@pq.cnpq.br}
\author{\ R.S. Perez}
\email{rperez@cbpf.br}
\affiliation{Departamento de F\'isica, UFES, Avenida Fernando Ferrari, 514, CEP 29075-910, Vit\'oria, ES, Brazil.}
\author{\ N. Pinto Neto}\email{nelson.pinto@pq.cnpq.br }
\affiliation{Centro Brasileiro de Pesquisas Fisicas, Rua
Xavier Sigaud, 150, CEP 22290-180, Rio de Janeiro, Brazil.
}
\author{\ Santiago E. Perez Bergliaffa}
\email{sepbergliaffa@gmail.com}
\affiliation{Departamento de F\'{\i}sica Te\'{o}rica,
Instituto de F\'{\i}sica, Universidade do Estado de Rio de
Janeiro, CEP 20550-013, Rio de Janeiro, Brazil.}
\vspace{.5cm}

\pacs{98.80.-k, 04.50.Kd}

\begin{abstract}

Several features of an $f(R)$ theory in which there is a maximum value for the curvature are analyzed. The theory
admits all the
vacuum solutions of General Relativity, and also the radiation evolution for the scale factor of the standard cosmological model.
Working in the Jordan frame, a complete analysis of the phase space is performed, and shown to be in agreement with
examples obtained by numerical integration. In particular, nonsingular cosmological solutions were obtained, which after the bounce enter a phase of de Sitter expansion and subsequently relax to a GR-like radiation-dominated evolution.

\end{abstract}

\maketitle

\section{Introduction}

Although General Relativity (GR) agrees with observation with good precision in several situations
\cite{will}, theories described by a Lagrangian that is a function of the Ricci and other scalars constructed with the Riemann tensor have been intensively studied in recent times. These theories, which are carefully engineered
in such a way that they differ from GR either in the limit of low or high curvature (or in both limits) have a long tradition, starting with a paper by Weyl in 1918 \cite{weyl}. The current interest
in these theories is twofold. First -in the case of low curvature- the aim is to model the
accelerated expansion of the universe that follows from several observations \cite{acc} (when
interpreted in the standard cosmological model \cite{kra})
without using dark energy \cite{capo}. Second, since we have yet no direct observational evidence of the behaviour of the gravitational field for high values of the curvature, these theories are used to widen our
understanding of possible new features of gravity in this regime.
Black holes \cite{perro} and neutron stars \cite{psal} are natural candidates to look for deviations from GR in the strong regime, as well as the universe in its phase of very high density.
We shall be interested here precisely in the latter circumstance, by restricting to theories with Lagrangian that is a function of the Ricci scalar only \cite{review}. This option is due to simplicity,
and to the fact that $f(R)$ theories are favoured over more general choices because they do not suffer from the
Ostrogradski instability \cite{ostro}.

The cosmology of $f(R)$ theories in the large-curvature limit has been explored for instance in \cite{staro}, where  inflation was analyzed in the theory $f(R) = R + R^2/M^2$. Still in this regime, $f(R)$ theories have been used to
remedy one of the main problems of the standard cosmological problem, namely the initial singularity \cite{bouncepr}. In this regard, general conditions for the absence of an initial singularity in $f(R)$ theories were discussed in \cite{bounce}, while solutions displaying a bounce for specific forms of $f(R)$ were exhibited in
\cite{kerner}.\\
In this article, we shall begin the analysis of the
large-curvature characteristics of the theory given by
\begin{equation}
f(R) = R - \beta \sqrt{1-\frac{R^2}{\beta^2}} + \beta,
\label{lag}
\end{equation}
where $\beta$ is a constant that explicitly sets un upper limit for the curvature, in a way reminiscent of that of
Born-Infeld's electromagnetism \cite{bi}. Notice however that our proposal is different from the
so-called Born-Infeld gravity, which has been discussed for instance in \cite{big}. The theory examined here is also different from the one studied in
\cite{schmidt}, with Lagrangian given by
\begin{equation}
{\cal L} = \frac{1}{16\pi G}\frac{R}{\sqrt{1 + l^4 R^2}},
\end{equation}
(where $l$ is a dimensional constant)
since in this theory the curvature scalar is not bounded and the effective gravitational constant $G_{\rm eff}$ (which is proportional to $\frac{\partial f}{\partial R}$) diverges for large $R$,
while in the theory defined by Eqn.(\ref{lag}) the curvature is limited by
the parameter $\beta$, and $G_{\rm eff}$  is finite for all $R$ .

Our main goal will be to characterize the cosmological solutions of the theory
given by Eqn.(\ref{lag}) in the strong field regime (in the presence of radiation), giving particular attention to solutions that do not display an initial singularity.
The general features of this theory, as well as its restriction to the standard cosmological model
will be discussed in Section \ref{gf}. A phase space analysis of the resulting equations will be presented in
Section \ref{psa} \cite{psanalysis}, while relevant examples of this analysis will be illustrated by numerical integration
in Section \ref{ni}. We close with a discussion of the results in Section \ref{con}.

\section{\label{gf} General features}

The starting point is the gravitational action given by
\begin{equation}
S_g = \frac{1}{2\kappa} \int d^4 x\sqrt{-g} f(R),  \label{action}
\end{equation}
where $\kappa = 8\pi G$, $G$ being the gravitational constant, we have set $c=1$, and $f(R)$ is a general function of the Ricci scalar. As mentioned above, we shall work with the function
\begin{equation}
f(R) = R - \beta \sqrt{1-\frac{R^2}{\beta^2}} + \beta. \label{fr}
\end{equation}
The form above is similar to the lagrangian proposed by Born and Infeld to solve the problem of singularities in the electromagnetic field, and so it carries their name. $\beta$ is a free parameter of the theory, hopefully to be identified with a fundamental constant of nature,
like the inverse of the Planck length squared.
This function bounds the value of the Ricci scalar, both from below and from above, with $|R|<\beta$, therefore singularities characterized by the divergence of $R$ should be supressed in the solutions of this theory. The third term is inserted to recover Minskowski spacetime in the absence of curvature.
The equations of motion that follow from the variation of the action (\ref{action}) with respect to the metric are,
\beq
\frac{df(R)}{dR}R_{\mu\nu}-\half f(R)g_{\mu\nu}-\left[\nabla_\mu\nabla_\nu-g_{\mu\nu}\Box\right]\frac{df(R)}{dR}=
\kappa T_{\mu\nu},
\label{eom4}
\eeq
where $T_{\mu\nu}$ is the energy-momentum of the matter fields, defined by
$$
T_{\mu\nu} = -\frac{2}{\sqrt{-g}}\frac{\delta S_{\rm M}}{\delta g^{\mu\nu}},
$$
and the covariant derivative
is defined using the usual Levi-Civita connection.
Taking the trace, we obtain
\beq
\frac{df(R)}{dR}R-2f(R)+3\Box\frac{df(R)}{dR}=\kappa T,
\label{trace1}
\eeq
We see from Eqns.(\ref{eom4}) and (\ref{trace1}) in the case of the $f(R)$ given by Eqn.(\ref{fr}) that all solutions of GR with a traceless energy-momentum tensor as a source are solutions of the $f(R)$ theory studied here, although
they may not be unique. The same can be said about the vacuum solutions of GR.

Regarding the two basic conditions that a given $f(R)$ should satisfy in order to be a viable theory, namely $df/dR>0$ and $d^2f/dR^2>0$ \cite{review}, a simple calculation shows that the second one
is satisfied for any value of $|R| <\beta$, while the first one is violated for $R<-0.71\beta$. However, considering that $\beta$ is of the order of
the inverse of Planck length squared, this violation takes place when the scalar curvature scale of spacetime,
defined as $l_c \equiv |R|^{-1/2}$,
approachs the Planck length, $l_c > l_{\rm Planck}$, where quantum gravitational effects migth be important and the theory
should be modified. We shall assume that the theory has plenty of solutions that do not reach this regime, and
show below that this
assumption is well-founded in a cosmological context.

Since $R$ contains second derivatives of the fundamental coordinate $g_{\mu\nu}$, the metric tensor, the field equations arising from it are of fourth order in the derivatives of $g_{\mu\nu}$. Both the analytical and numerical handling of the resulting equations
is easier if we
introduce the auxiliary field $s \equiv R$, in such a way that the action is now written as
\begin{equation}
 S_g = \frac{1}{2\kappa} \int d^4 x \sqrt{-g} \left[ f(s) - \phi (s-R) \right], \label{jordan}
\end{equation}
where $\phi$ is a Lagrange multiplier.
Variation of the action (\ref{jordan}) with respect to $s$ yields $\phi = f'(s)$, where the prime denotes derivative with respect to the argument. The equation (\ref{jordan}) can be rewritten as
\begin{equation}
S_g = \frac{1}{2\kappa} \int d^4 x \sqrt{-g} \left[ \phi R - V(\phi) \right] , ~~~~ V(\phi) = \phi  \, s(\phi) - f[s(\phi)], \label{bransdicke}
\end{equation}
so  the $f(R)$ theory is identical to a Brans-Dicke theory with parameter $\omega=0$. The relation between $\phi$ and $s$ is as follows:
\begin{equation}
\phi(s)=f'(s)= 1 + \frac{s}{\sqrt{\beta^2-s^2}} ~~~~~\rightarrow ~~~~~~   s(\phi) = \pm \beta \sqrt{\frac{(\phi-1)^2}{1+(\phi-1)^2}}
\end{equation}

The variation of (\ref{bransdicke}) with respect to $\phi$ results in
\begin{equation}
R = V'(\phi), \label{RV}
\end{equation}
It is possible to see that, in order to be consistent with Eq.~(\ref{RV}) one must set,
\begin{equation}
s(\phi) = \beta \frac{\phi-1}{\sqrt{1+(\phi-1)^2}} .
\end{equation}
Hence, the potential must have the form
\begin{equation}
 V(\phi) = \beta \left( \sqrt{1+(\phi-1)^2} -1 \right)
\end{equation}
In the situation of zero curvature, $\phi=1$ and $V(\phi)=0$, and Minkowski spacetime is regained. Finally, the variation of the action (\ref{bransdicke}) with respect to the metric tensor. We obtain the following field equations
\begin{equation}
\phi G_{\mu\nu} + \frac{1}{2} V(\phi) g_{\mu\nu} - \nabla_\mu \nabla_\nu \phi + g_{\mu\nu} \Box \phi = \kappa T_{\mu\nu}, \label{field}
\end{equation}
and the associated trace is
\begin{equation}
 \Box \phi - \frac{1}{3} \phi V'(\phi) + \frac{2}{3} V(\phi) = \frac{\kappa}{3} T, \label{trace}
\end{equation}
where we used Eqn.(\ref{RV}). Matter has been included in the right hand side of these equations, as the energy-momentum tensor $T_{\mu\nu}$, with trace $T$.

\begin{figure}[ht]
\begin{center}
\includegraphics[height=6cm]{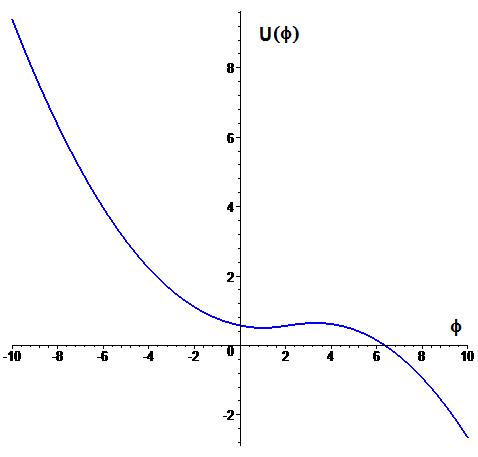}
\caption{The effective potential $U(\phi)$ displays a local minimum at $\phi=1$, corresponding to GR, and a local maximum at $\phi=3.315$.}
\label{graphphi}
\end{center}
\end{figure}

Armed with these equations, we turn next to their action on the geometry of the standard cosmological model.
The background evolution is obtained by inserting the Friedman-Lem\^aitre-Robertson-Walker (FLRW) metric in Eqns.(\ref{field}) and (\ref{trace}), and considering a perfect fluid form for the energy-momentum tensor. Then, we get for the $00$ and $11$ components and for the trace, respectively,
\begin{equation}
H^2 + H \dot{\phi} = \frac{V}{6} + \frac{\rho}{3}
\label{ds1}
\end{equation}
\begin{equation}
(2 \dot{H} + 3 H^2)\phi = \frac{V}{2} - 2H\dot{\phi} - \ddot{\phi} - p
\label{ds2}
\end{equation}
\begin{equation}
\ddot{\phi} + 3 H \dot{\phi} + \frac{1}{3} \phi V' - \frac{2}{3} V = \frac{1}{3} (\rho - 3p)
\label{ds3}
\end{equation}
Since we are interested in the early universe, which was radiation dominated,
the right-hand-side of the trace equation is zero, and we can write it as the Klein-Gordon equation for the field $\phi$
\begin{equation}
 \ddot{\phi} + 3 H \dot{\phi} + U'(\phi) = 0, \label{kleingordon}
\end{equation}
with the effective potential
\begin{equation}
U(\phi) = -\frac{\beta}{6} \left[ \phi \sqrt{1+(\phi-1)^2} - 3 \sqrt{1+(\phi-1)^2} + 3\, {\rm arcsinh}(\phi-1) - 4 \phi \right]
\label{effpot}
\end{equation}
shown in Fig.\ref{graphphi}. This plot justifies \textit{a posteriori} the use of the so-called Jordan frame, since the potential is not multivalued.

The system of coupled equations (\ref{ds1})-(\ref{ds3}) is very complicated to be solved analytically. The plot shown in Fig. \ref{graphphi} will be important to qualitatively understand the different behaviors that will be obtained when we numerically solve our system of equations. We will see that the local maximum and minimum in the potential correspond to critical points of the dynamical system associated the evolution equations.

\section{\label{psa} Dynamical system analysis}

The dynamical system is built from Eqns.(\ref{ds1})-(\ref{ds3}) and (\ref{kleingordon}) and is given by
\begin{eqnarray}
\dot{H} &=& -2H^2 + \frac{V'}{6},  \label{dyn1} \\
\dot{y} &=& -3Hy - \frac{1}{3} \phi V' + \frac{2}{3} V , \label{dyn2} \\
\dot{\phi} &=& y .\label{dyn3}
\end{eqnarray}
Here the curvature scalar $R$ has been expressed in terms of the scale factor and its derivatives, taking the Hubble parameter as the fundamental function, and we introduced the variable $y$ so that the Klein-Gordon equation can be reduced to two first order equations.

\subsection{Critical points}

The critical points, which correspond to those points where
$\dot{H}=\dot{y}=\dot{\phi}=0$, are given by
\begin{equation}
H=\pm\sqrt{V'/12}, ~~~~~ \phi V' = 2 V, ~~~~~~ y=0.
\end{equation}
From the relation $\phi V' = 2 V$, we find the following equation values for $\phi$:
\begin{eqnarray}
\phi &=& 1,\\
u^3 - 2 u^2 +u - 4 &=& 0, ~~~~ (u=\phi-1).
\end{eqnarray}
So we have in principle four values of $\phi$, leading to 8 critical points if we consider $H=\pm \sqrt{V'/12}$. But the third order polynomial equation has only one real root, and so we have in principle four critical points. Also, $\phi = 1$ corresponds to $H = V = V' = 0$. Hence we end up with three critical points given by:
\begin{eqnarray}
P_1  & \rightarrow & \phi=1,~ H=0, ~ y=0; \label{p1} \\
P_2  & \rightarrow & \phi=3.315, ~ H=+\sqrt{V'/12},~ y=0; \label{p2}\\
P_3  & \rightarrow & \phi=3.315,~ H=-\sqrt{V'/12}, ~ y=0.  \label{p3}
\end{eqnarray}
The first point, $P_1$, corresponds to the Minkowski spacetime, while $P_2$ to a expanding de Sitter and $P_3$ to a contracting de Sitter,
both with a constant scalar curvature $R = 0.92 \beta$. Notice that the values obtained for $\phi$ correspond to the local minimum and maximum in Fig. \ref{graphphi}. To find the nature of these critical points we must linearize the system (\ref{dyn1})-(\ref{dyn3}), by replacing
\begin{equation}
H = H_0 + \delta H, ~~~~~~ \phi = \phi_0 + \delta \phi, ~~~~~ y= \delta y,
\end{equation}
where $H_0$ and $\phi_0$ are the values of $H$ and $\phi$ at the critical points (notice that $y_0=0$ for all the critical points), given by
Eqns.(\ref{p1})-(\ref{p3}). We then find the following linearized system
\begin{eqnarray}
\delta \dot{H} &=& - 4H \, \delta H + \frac{1}{6} V'' \delta\phi ,\\
\delta \dot{y} &=& -3H \, \delta y - 3y \, \delta H - \frac{1}{3} V' \delta\phi - \frac{1}{3} \phi V'' \delta\phi + \frac{2}{3}V'\delta\phi ,\\
\delta\dot{\phi} &=& \delta y,
\end{eqnarray}
where we have omited the subindex 0 on the background quantities.
We shall investigate next each point separately.

\begin{itemize}
\item Critical point $P_1$\\
Corresponding to Minskowski spacetime, this point has $H_0=0$ and $\phi_0=1$. The equations around this point are
\begin{eqnarray}
\delta\dot{H} &=& \frac{\beta}{6} \delta\phi,  \\
\delta\dot{y} &=& - \frac{\beta}{3} \delta\phi, \\
\delta\dot{\phi} &=& \delta y,
\end{eqnarray}
which in matrix form read
\begin{equation}
\left( \begin{array}{c}
\delta\dot{H}\\
\delta\dot{y}\\
\delta\dot{\phi}
\end{array} \right) = \left( \begin{array}{ccc}
0 & 0 & \frac{\beta}{6} \\
0 & 0 & -\frac{\beta}{3} \\
0 & 1 & 0 \end{array} \right) \left( \begin{array}{c}
\delta H \\ \delta y \\ \delta\phi \end{array} \right)
\end{equation}
The diagonalisation of this system involves the calculation of the eigenvalues $\lambda$, given by
\begin{equation}
{\rm det} \left( \begin{array}{ccc}
-\lambda & 0 & \frac{\beta}{6}\\
0 & -\lambda & -\frac{\beta}{3}\\
0 & 1 & -\lambda    \end{array} \right) = -\lambda^3 - \frac{\beta}{3} \lambda = 0 .
\end{equation}
The roots are
\begin{equation}
\lambda = \pm i \sqrt{\frac{\beta}{3}}, ~~~~ \lambda = 0.
\end{equation}
Since one eigenvalue is zero and the other two are pure imaginary with opposite signs, this critical point is a center-saddle point.

\item Critical points $P_2$ and $P_3$\\
These are de Sitter spacetimes. Following the same approach used for $P_1$, the linearized equations around the points $P_2$ and $P_3$ are
\begin{eqnarray}
\delta \dot{H} &=& -4 H \, \delta H + \frac{1}{6} V'' \delta\phi\\
\delta \dot{y} &=& - 3H \, \delta y + \frac{1}{3}V' \delta \phi - \frac{1}{3} \phi V'' \delta\phi\\
\delta \dot{\phi} &=& \delta y
\end{eqnarray}
To obtain the eigenvalues, we must solve the equation
\begin{equation}
\lambda^3 + 7H \lambda^2+ \left( \frac{2}{3}V' + \frac{1}{3} \phi V'' \right)\lambda - \frac{4}{3} H(V'-\phi V'')=0
\end{equation}
For the critical point $P_2$, we have $\phi=3.3146$ and $H = \sqrt{V'/12}$, therefore
\begin{equation}
\lambda_1 = -1.10634,~~~ \lambda_2 = -1.05456,~~~ \lambda_3 = 0.224804.
\end{equation}
Since we have two eigenvalues with negative sign and one with positive sign, this indicates that it is a saddle point in two different planes, while it is a attractor in another.

For $P_3$ we put $\phi=3.3146$ and $H = -\sqrt{V'/12}$, yielding
\begin{equation}
\lambda_1 = -0.224804,~~~  \lambda_2 = 1.05456, ~~~ \lambda_3 = 1.10634.
\end{equation}
Once again, we have a saddle point in two different planes, but now it is an repeller in the third plane. We cannot tell which planes are those for now, but we will see that the repulsive/attractive behavior of these de Sitter points will be present in the numerical evolutions.

\end{itemize}

\subsection{Critical points at infinity}

The complete phase diagram can be determined from the knowledge of the
critical points at infinity. Let us write a three-dimensional system as follows:
\begin{eqnarray}
 \dot{x} &=& X(x,y,z),\\
\dot{y} &=& Y(x,y,z),\\
\dot{z} &=& Z(x,y,z),
\end{eqnarray}
and define the new variables
\begin{equation}
x = \frac{u}{k}, ~~~ y=\frac{v}{k}, ~~~ z=\frac{w}{k},
\end{equation}
with the constraint
\begin{equation}
u^2 + v^2 + w^2 + k^2 = 1
\end{equation}
The new variables define a three-sphere, with $k = 0$ denoting the ``equator" of the three-sphere, in fact a two-sphere. The two-sphere defined in this way corresponds to the infinity. Hence, we have now
\begin{equation}
\frac{\dot{u}}{k} - \frac{\dot{k}}{k^2} u  = X (u/k,v/k,w/k)
\end{equation}
\begin{equation}
\frac{\dot{v}}{k} - \frac{\dot{k}}{k^2} v = Y(u/k,v/k,w/k)
\end{equation}
\begin{equation}
\frac{\dot{w}}{k} - \frac{\dot{k}}{k^2} w = Z(u/k,v/k,w/k)
\end{equation}
with the condition
\begin{equation}
 u\, du + v\, dv + w\, dw + k\, dk = 0
\end{equation}
Combining these equations we obtain the following system:
\begin{eqnarray}
\dot{u} & = & (1-u^2)kX - kuwZ -kuvY\\
\dot{v} &=& (1-v^2) kY - kvwZ - kuvX\\
\dot{w} &=& (1-w^2)kZ - kuwX - kvwY
\end{eqnarray}
In our specific case, putting $H \rightarrow x$, $\phi \rightarrow z$ and keeping $y$ as it is, we have the expressions
\begin{equation}
X = \frac{1}{k^2} \left\{ -2 u^2 + \frac{\beta}{6} \frac{k^2 (w-k)}{\sqrt{k^2+(w-k)^2}} \right\} = \frac{1}{k^2} \bar{X}
\end{equation}
\begin{equation}
Y = \frac{1}{k^2} \left\{ -3uv -\frac{\beta}{3}\frac{kw(w-k)}{\sqrt{k^2+(w-k)^2}} + \frac{2\beta k}{3} \left[\sqrt{k^2+(w-k)^2} -1  \right] \right\} = \frac{1}{k^2} \bar{Y}
\end{equation}
\begin{equation}
Z= \frac{1}{k^2} kv = \frac{1}{k^2} \bar{Z}
\end{equation}
Using the expressions for our specific system, redefining the evolutive parameter, such that $d\tau / k \rightarrow dt$, we end up with the following dynamical system at infinity ($k = 0$):
\begin{eqnarray}
\dot{u} &=& -2 u^2 (1-u^2) + 3 u^2 v^2\\
\dot{v} &=& - 3uv(1-v^2) + 2u^3 v\\
\dot{w} &=& 2u^3 w + 3 u v^2 w
\end{eqnarray}
There are the following critical points:
\begin{itemize}
\item $P_4$: $u = v = w = 0$. Incompatible with the condition (at infinity) $u^2 + v^2 + w^2=1$.

\item $P_5$: $u = \pm 1, v = 0, w = 0$. This critical point is located on the top and on the
bottom of the $u$ axis, that is, the axis of $H$ at infinity, with the scalar field and its deriviative equal to zero.

\item $P_6$: $u = 0$, for all $v,w$ satisfying the condition $v^2 + w^2 = 1$, or $\dot{\phi}^2 + \phi ^2 \to \infty$.
\end{itemize}

The critical point $P_5$ represents the beginning of the universe from a singularity (the big bang), sign plus, or the end of the universe in a singularity (big crunch), sign minus, the time reversal of the first case.

The critical points $P_6$ lying on the sphere $v^2 + w^2 = 1$ , or $\dot{\phi}^2 + \phi ^2 \to \infty$,
correspond to the physical situation where $\phi$ is deep down in the potential $U(\phi)$ given in
Eq.~(\ref{effpot}). Looking at Eq.~(\ref{dyn1}), this corresponds to contracting or expanding de Sitter spacetimes,
where the Ricci scalar is saturated to its maximum constant value, $R=\beta$ (that is the reason for having $u = kH = k\sqrt{12 R} = 0$
when $k=0$).

Linearizing the system around
the critical point $P_5$, we obtain, for that
\begin{equation}
 \delta \dot{u} = \pm 4 \delta u .
\end{equation}
This implies that the critical point $u = 1$ is a repeller (the curve emerges from
this point) and $u = -1$ is an attractor: the first
critical point is the big bang while the second is the big crunch.

Linearization of the equations around the critical point $P_6$,
leads to the following system:
\begin{eqnarray}
\delta\dot{u} &=& 0,\\
\delta\dot{v} &=& -3(1-v^2) \delta u,\\
\delta\dot{w} &=& 3v^2 w \, \delta u.
\end{eqnarray}
The first of these equations imply that $\delta u$ remains constant. Using the fact that $u^2 + w^2 = 1$, we obtain the system
\begin{eqnarray}
\delta \dot{v} &=& - 3 v w^2 \delta u\\
\delta \dot{w} &=& 3 v^2 w \, \delta u
\end{eqnarray}
This implies the following configurations:
\begin{itemize}
\item $\delta u > 0$\\
- $v > 0$, $w > 0$ implies that the trajectories are repulsive in the direction $w$ and attractive in the direction $v$;\\
- $v > 0$, $w < 0$ implies that the trajectories are attractive in both the directions $v$ and $w$;\\
- $v < 0$, $w > 0$ implies that the trajectories are repulsive in both the directions $v$ and $w$;\\
- $v < 0$, $w < 0$ implies that the trajectories are attractive in the direction $w$ and repulsive in the direction $v$.

\item $\delta u < 0$:\\
- $v > 0$, $w > 0$ implies that the trajectories are attractive in the direction $w$ and repulsive in the direction $v$;\\
- $v > 0$, $w < 0$ implies that the trajectories are repulsive in both the directions $v$ and $w$;\\
- $v < 0$, $w > 0$ implies that the trajectories are attractive in both the directions $v$ and $w$;\\
- $v < 0$, $w < 0$ implies that the trajectories are repulsive in the direction $w$ and attractive in the direction $v$.
\end{itemize}

After this analysis, one can envisage many scenarios. The less interesting one is the usual radiation-dominated, spatially-flat FLRW solution, which
is allowed in this framework, coming from the critical point $P_5$, $u = 1$ (the big bang singularity) and going to the critical point $P_1$, $H = 0$ (Minkowski), always with $\phi = 1$. However, one may have many other singular solutions, e.g., one coming from the singular point, $P_5$, $u = 1$, but away from the line $\phi = 1$ at infinity, which may cross the plane defined by $H = 0$, and go to the critical point at $\phi \rightarrow -\infty$ by below.

The region that may exhibit nonsingular solutions is the one with positive $\phi$. Some of them can begin at the critical point $P_1$, contracting from Minkowski spacetime to the critical point $P_3$ (contracting de Sitter spacetime), then going to the critical point $P_2$ (expanding de Sitter spacetime), with a bounce in between, where the plane $H=0$ is crossed.
The trajectory then returns to the point $P_1$, that is, expanding to Minkowski spacetime.
The scalar field starts to oscillate very close to $\phi=1$ at $t\to -\infty$  (the general relativistic limit), goes to the local maximum of the potential (due to anti-friction because of contraction, see Fig.\ref{graphphi}), and then returns to the local minimum at $\phi=1$ at $t\to -\infty$. This is a nonsingular
solution with a bounce and an almost de Sitter expansion happening before a decelerated expansion phase, close to the usual radiation-dominated, spatially-flat FLRW general relativistic solution.

Other solutions are those contracting from the singular point $P_6$, a de Sitter spacetime with $R\approx\beta$ and $\phi \to\infty$ at $t\to -\infty$, in which the scalar field climbs the potential from below, due to anti-friction.
The solution crosses the plane $H=0$, thus going through a bounce, and reaches the critical $P_2$, near the local maximum of the potential (see Fig.\ref{graphphi}),
and then moves to the critical point $P_1$, with $\phi \approx 1$ at $t\to\infty$. Again, one has a non singular
solution with a bounce and an almost de Sitter expansion happening before a deccelerated expansion phase close to the
radiation-dominated, spatially-flat FLRW general relativistic
solution. The difference is that the former solution corresponds to a nonsingular bouncing
solution between two Minkowski spacetimes, while the second corresponds to a
nonsingular bouncing
solution between de Sitter contraction and Minkowski spacetime. Such trajectories are presented in the next section.

\section{\label{ni} Numerical Results}

The dynamical system (\ref{dyn1})-(\ref{dyn3}) can be numerically solved by simple inputs in programs like Maple or Mathematica. However, this system is highly nonlinear and thus very sensitive to initial conditions. We will begin our evolution around different critical points and see how the trajectories behave for small fluctuations ($\delta H$, $\delta\phi$ and $\delta y$). For numerical considerations, we adopt $\beta = 0.5$.

From Fig.\ref{graphphi}, we see that the critical points of the finite region are the local minimum (Minkowski configuration) and maximum (de Sitter configuration). We will now list interesting examples obtained from particular sets of $H$, $\phi$, $y$ and their fluctuations.

\begin{itemize}
\item Around critical point $P_1$\\
Starting from a Minkowski spacetime, the value of $\phi$ begins to oscillate with increasing amplitude, while $H$ oscillates around a decreasing average value. Eventually, $\phi$ overcomes the local maximum at its right and rolls down the potential, towards $\phi \rightarrow +\infty$. This leads to a big crunch singularity, with $H \rightarrow -\infty$. The evolution is shown below.

\begin{figure}[ht]
\begin{center}
\includegraphics[height=6cm]{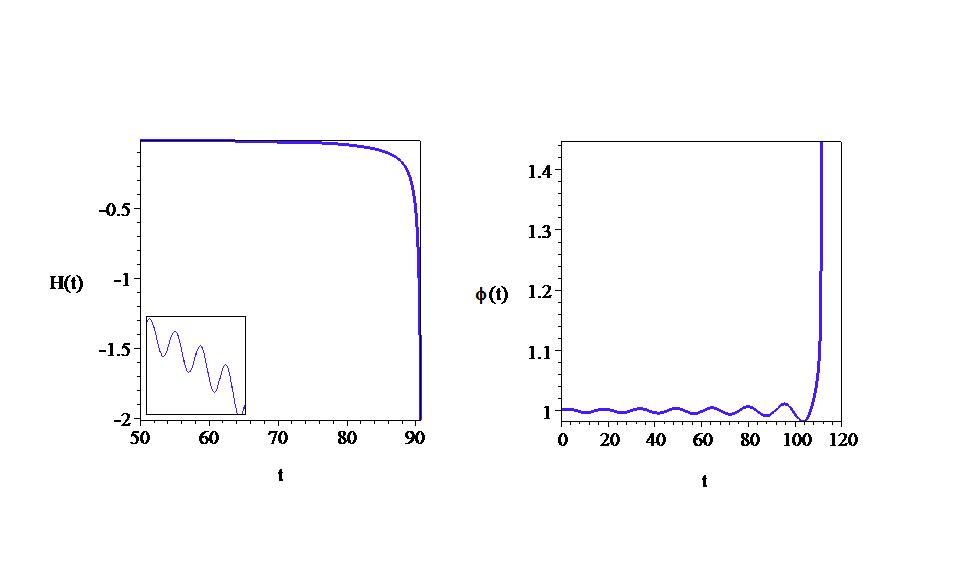}
\caption{Evolution for $H$ and $\phi$ around Minkoski spacetime. The detailed frame shows the oscillation of the Hubble parameter between $t=-80$ to $t=-20$.}
\end{center}
\end{figure}

\item Around critical point $P_2$\\
Starting our evolution around the local maximum is very tricky, since fine tunning of $\delta\phi$ and $\delta y$ is needed in so that the field does not roll down the potential to $\phi \rightarrow + \infty$. An interesting solution is shown below (Fig.\ref{hphi4}), representing an universe emerging from a past singularity, as in the big bang model, and reaching the Minkowski configuration after an intermediate de Sitter phase, which is a saddle point. This is precisely the evolution presented in the cosmological Standard Model.

\begin{figure}[ht]
\begin{center}
\includegraphics[height=6cm]{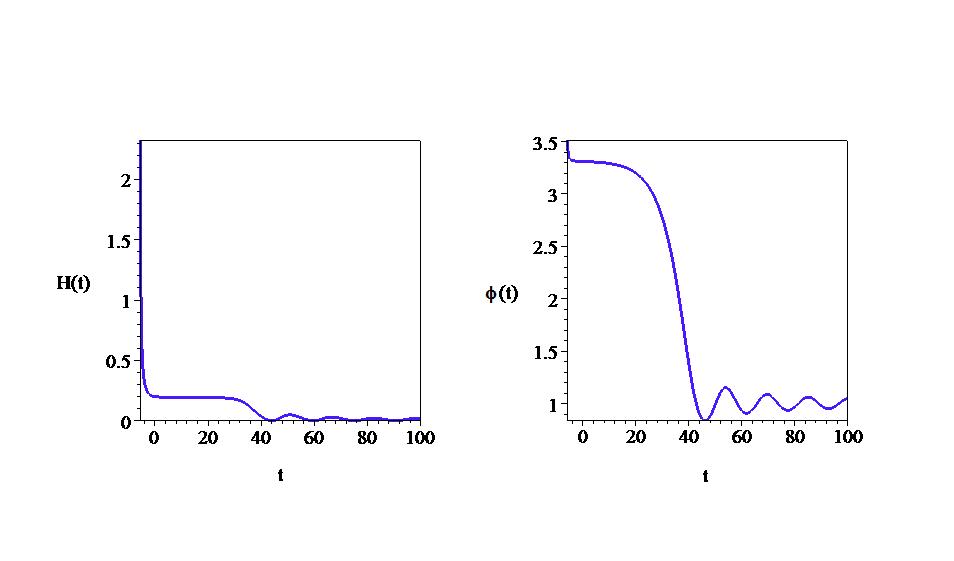}
\caption{\label{hphi4} Evolution of a universe that emerges from a past singularity, enters a
de Sitter phase and reaches Minkowski spacetime.}
\end{center}
\end{figure}

\item Around critical point $P_3$\\
In this case, we have the same situation as in $P_2$, with the substitution $H \rightarrow -H$ and $t \rightarrow -t$. The universe begins in a Minkowski state, but the Hubble parameter begins to oscillate. Then, it goes through a contracting de Sitter phase which culminates in a big crunch singularity, with $H \rightarrow -\infty$

\begin{figure}[ht]
\begin{center}
\includegraphics[height=6cm]{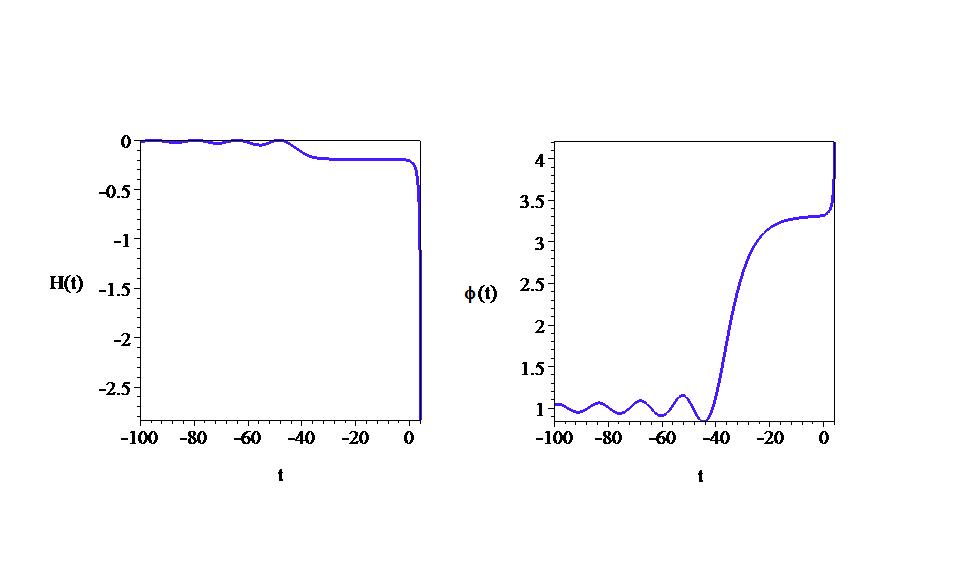}
\label{hphi6}
\caption{In this case the Hubble parameter oscillates around the Minkowski state, and goes through a contracting de Sitter phase which culminates in a big crunch singularity.}
\end{center}
\end{figure}

\item Bouncing solutions

\begin{itemize}

\item Minkowski-to-Minkowski evolution: taking the starting point of our evolution as the intermediate point between both de Sitter configurations, with $\phi \approx 3.315$ and $H \approx 0$ we obtain trajectories characterizing nonsingular universes. Two examples are
shown below. The first one begins in a Minkowski state which oscillates with growing amplitude, until it reaches a contracting de Sitter phase. Suddenly, the universe bounces to a de Sitter expansion, which lasts for a while. Then, it goes through oscillations once again to finally reach a Minkowski configuration
(see Fig.\ref{bounce1}).  This is the only solution entirely contained in the finite region, with $\phi$ always to the left of the local maximum of the effective potential shown in Fig.\ref{graphphi}.

\begin{figure}[ht]
\begin{center}
\includegraphics[height=7cm]{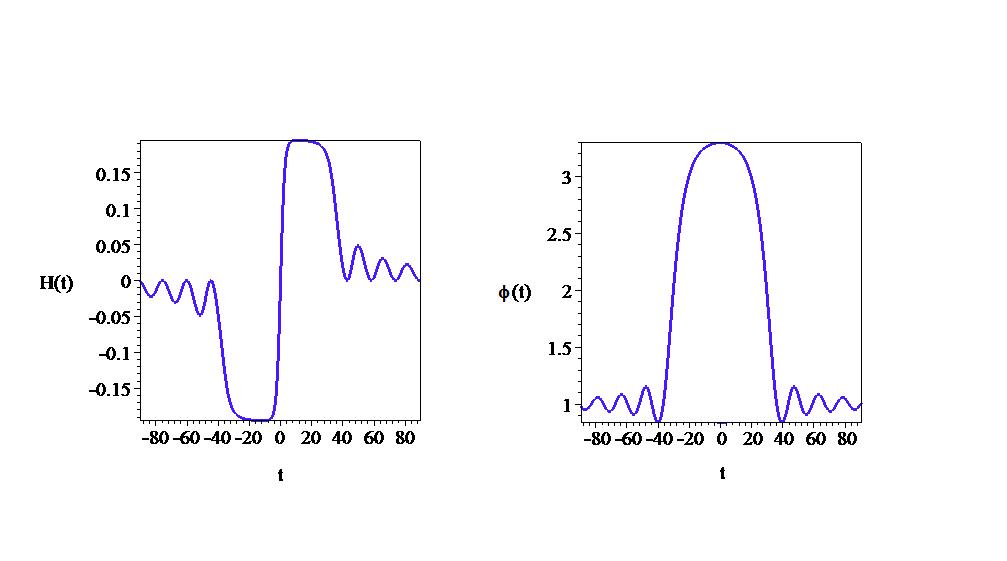}
\caption{\label{bounce1}
The model in the figure begins with growing oscillations around the Minkowski state, followed by a contracting de Sitter phase, a bounce to a de Sitter expansion, to finally reach again the Minkowski geometry.}
\end{center}
\end{figure}

\item de Sitter-to-Minkowski evolution: the second situation (Fig.\ref{bounce2})
also presents a bounce, but it starts from a contracting de Sitter state at $t \rightarrow - \infty$. This reflects the result we obtained from the analysis of the critical point $P_3$, which behaves as a repeller in this particular plane.

\begin{figure}[ht]
\begin{center}
\includegraphics[height=6cm]{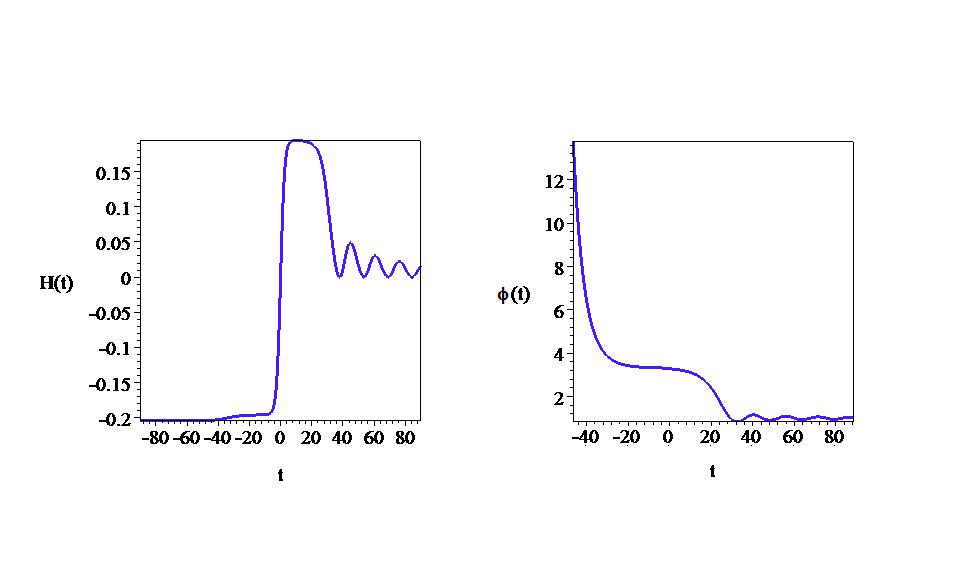}
\caption{\label{bounce2} In this case the evolution, having started from de Sitter spacetime, displays a bounce,  and reaches Minkowski in the end.}
\end{center}
\end{figure}

\item de Sitter-to-de Sitter evolution: once again, we get the repulsive and attractive behaviors of the de Sitter critical points in particular trajectories. The universe comes from an infinite de Sitter contraction in the past and goes through a bounce, acquiring afterwards a forever expanding de Sitter configuration (Fig.\ref{bounce3}). The field $\phi$ in this evolution is restricted to the right of the local maximum of the effective potential, i.e., $\phi > 3.315$ for all times.

\begin{figure}[ht]
\begin{center}
\includegraphics[height=6cm]{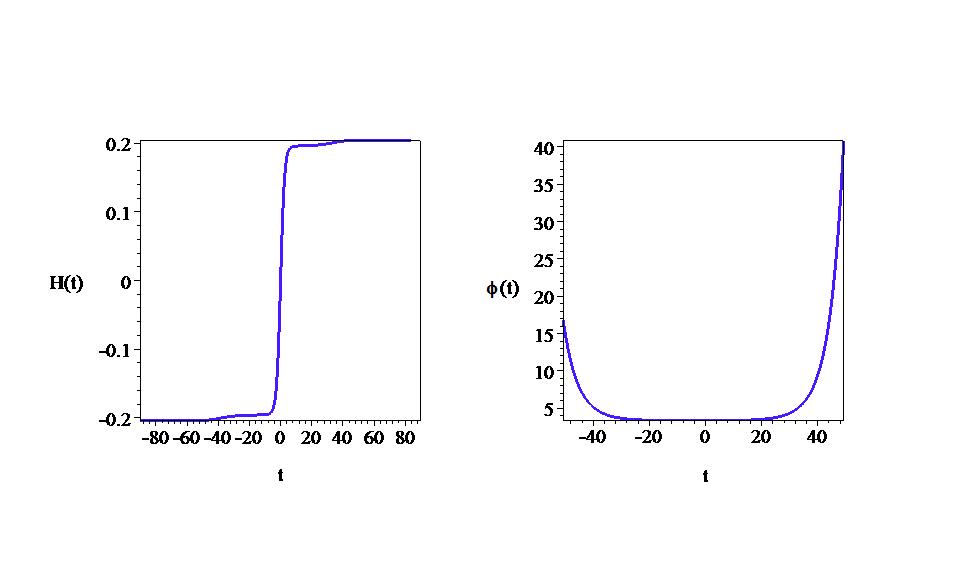}
\caption{\label{bounce3} The figure shows a de Sitter to de Sitter evolution
through a bounce.}
\end{center}
\end{figure}

\end{itemize}

A bouncing solution can also be obtained analytically by assuming that the Ricci scalar approaches its maximum possible value, given by $\beta$, at the bounce ($t=0$), so we may write
\begin{equation}
 R = \beta - \epsilon, ~~~~~ 0 < \epsilon << 1.
\end{equation}
Since $\epsilon$ is very small, the component 00 of Eqn.(\ref{eom4}) in a FLRW universe is now given by
\begin{equation}
 \frac{3\beta}{\sqrt{2\beta\epsilon}} \left[ \frac{\ddot{a}}{a} - \frac{\dot{a}\dot{\epsilon}}{2a\epsilon} \right] \simeq \kappa \rho,
\end{equation}
where we kept only the dominant terms, proportional to $1/\sqrt{\epsilon}$. Multiplying both sides by $\sqrt{2\beta\epsilon}$, and assuming that the energy density at the bounce is high but finite, we get
\begin{equation}
 \frac{\ddot{a}}{a} - \frac{\dot{a}\dot{\epsilon}}{2a\epsilon} \propto \sqrt{\epsilon}. \label{aepsilon}
\end{equation}
Expanding the scale factor in Taylor series around $t=0$, and imposing the conditions $\ddot{a}(t=0)>0$ and $\dot{a}(t=0)=0$ valid at the bounce, the form of the scale factor must be $a(t) = a_0 + a_2 t^2 + a_3 t^3$. Hence, the Ricci scalar is
\begin{equation}
 R = 6 \left(\frac{\ddot{a}}{a} + \frac{\dot{a}^2}{a^2}  \right) \simeq  12a_2 + 36 a_3 t,
\end{equation}
where we neglected orders higher than the linear, and we can identify $\beta = 12 a_2$ and $\epsilon = -36 a_3 t$, with $a_3 < 0$. However, inserting the scale factor in Eqn.(\ref{aepsilon}) we obtain
\begin{equation}
 \frac{\ddot{a}}{a} - \frac{\dot{a}\dot{\epsilon}}{2a\epsilon} = a_2 + 6 a_3 t,
\end{equation}
which is always different from zero and therefore our choice of $a(t)$ cannot be a solution of the field equations.

Although the presence of the term $a_3 t^3$ is inconsistent with equation (\ref{aepsilon}), the same does not happen with a solution of the form $a(t)= a_0 + a_2 t^2 + a_4 t^4$. The Ricci scalar in this case is
\begin{equation}
 R \simeq 6 \left[ 2a_2 + (2a_2^2 + 12a_4) t^2  \right] ,
\end{equation}
where we ignored terms of order higher than $t^2$. It follows that $\beta = 12a_2$ and $\epsilon = -12 (a_2^2 + 6 a_4) t^2$. Since $\epsilon >0$, we must impose $a_4 < - a^2 / 6$. Inserting these quantities in Eqn.(\ref{aepsilon}), we get
\begin{equation}
\frac{\ddot{a}}{a} - \frac{\dot{a}\dot{\epsilon}}{2a\epsilon} \simeq 2a_2 - \frac{4a_2 t^2}{2t^2} = 0,
\end{equation}
and so the scale factor expansion is indeed a solution. Extending the Taylor series to order $t^5$, in the form $a(t) = a_0 + a_2 t^2 + a_4 t^4 + a_5 t^5$, it is straightforward to show that $\epsilon = -12 (a_2^2 + 6a_4) t^2 + 120 a_5 t^3$, yielding
\begin{equation}
\frac{\ddot{a}}{a} - \frac{\dot{a}\dot{\epsilon}}{2a\epsilon} \propto t \propto \sqrt{\epsilon},
\end{equation}
and therefore (\ref{aepsilon}) is satisfied. We can also see from the Taylor expansion that both $\ddot{a}$ and $\dot{a}$ are finite at $t=0$, so the Ricci scalar is well behaved in this scenario. We conclude that bouncing solutions can be obtained from the Lagrangian (\ref{fr}) if the coefficient $a_3$ in the Taylor expansion of the scale factor around $t=0$ is null.

\end{itemize}

\section{\label{con}Final remarks}

We have discussed here several features of an $f(R)$ theory in which there is a maximum value for the curvature (in a way reminiscent of Born-Infeld electrodynamics). The theory, defined by the Lagrangian given in Eqn.(\ref{fr}),
admits all the vacuum solutions of GR, and also the radiation evolution for the scale factor of the standard cosmological model. Working in the Jordan representation, a complete analysis of the phase space was presented,
the main results of which were illustrated with examples obtained by numerical integration.
It was shown that there are cosmological solutions that attain a minimum value of the scale factor and then expand again, instead of displaying the so-called initial singularity. Perhaps the most interesting ones are those shown in
Figs.(\ref{bounce1}) and (\ref{bounce2}) in which after the bounce there is a phase of de Sitter expansion and a subsequent relaxation to GR, with the field $\phi$ oscillating around $\phi = 1$. In particular, the plots in Fig.(\ref{bounce1}) suggest that after the de Sitter expansion, the universe goes through a radiation-dominated evolution, at least for some time, before oscillating around the Minkowski stage.
\footnote{Notice that the main goal of our model is to furnish a dynamical behaviour different than that of GR in the large-curvature regime. A soft transition to more recent eras, such as those of dust or acelerated expansion, may be implemented by adding to our model terms that are important only for
low values of the curvature, a procedure that was used for instance in
\cite{capo2}. By construction, these terms will not spoil our results
since they are irrelevant at high curvatures.}

Although we worked in the Jordan frame, it is worth pointing out that we expect our main results to be valid also in the Einstein frame, whose metric is related to that of the Jordan frame
via the transformation $\tilde g_{\mu\nu} = \phi g_{\mu\nu}$. The relation
between the Hubble parameter in the two frames is given by
$$\tilde H = H + \frac 1 2 \frac{\dot \phi}{\phi^{3/2}}.$$
Since $H$, $\phi$ and its derivative (displayed in the
plots) are well-behaved functions, it follows
that the nonsingular models we obtained in the Jordan frame (all of which satisfy the condition $\phi > 0 $ for all values of $t$),
will also be nonsingular in the Einstein frame.
Moreover, since the conformal transformation is always regular, the instabilities due to transition from attractive and repulsive gravity \cite{starobinski} are absent in these bouncing models.

Future work includes the study of the dynamical system with the addition of non-relativistic matter and dark energy
\cite{ame},
and the investigation of the features of compact objects in this theory, such as neutron stars (along the lines presented in
\cite{sal}) and black holes.

\section*{Acknowledgments}
All the authors would like to thank CNPQ for financial support.
SEPB would also like to thank FAPERJ and UERJ for support.

\end{document}